\documentclass[a4paper,twocolumn,english,showpacs,aps,pra,fleqn,superscriptaddress]{revtex4-1}

\usepackage[T1]{fontenc}
\usepackage[utf8]{luainputenc}
\usepackage{amsmath}
\usepackage{amssymb}
\usepackage{graphicx}
\usepackage{hyperref}
\usepackage{babel}
\usepackage{wasysym}
\usepackage{microtype}
\usepackage{color}

\begin{document}

\title{Production of large $^{41}$K Bose-Einstein condensates using D1 gray molasses}

\author{Hao-Ze Chen}
\affiliation{Shanghai Branch, National Laboratory for Physical Sciences at Microscale and Department of Modern Physics, University of Science and Technology of China, Hefei, Anhui 230026, China}
\affiliation{CAS Center for Excellence and Synergetic Innovation Center in Quantum Information and Quantum Physics, University of Science and Technology of China, Shanghai, 201315, China}
\affiliation{CAS-Alibaba Quantum Computing Laboratory, Shanghai, 201315, China}

\author{Xing-Can Yao}
\affiliation{Shanghai Branch, National Laboratory for Physical Sciences at Microscale and Department of Modern Physics, University of Science and Technology of China, Hefei, Anhui 230026, China}
\affiliation{CAS Center for Excellence and Synergetic Innovation Center in Quantum Information and Quantum Physics, University of Science and Technology of China, Shanghai, 201315, China}
\affiliation{CAS-Alibaba Quantum Computing Laboratory, Shanghai, 201315, China}
\affiliation{Physikalisches Institut, Ruprecht-Karls-Universität Heidelberg, Im Neuenheimer Feld 226, 69120 Heidelberg, Germany}

\author{Yu-Ping Wu}
\affiliation{Shanghai Branch, National Laboratory for Physical Sciences at Microscale and Department of Modern Physics, University of Science and Technology of China, Hefei, Anhui 230026, China}
\affiliation{CAS Center for Excellence and Synergetic Innovation Center in Quantum Information and Quantum Physics, University of Science and Technology of China, Shanghai, 201315, China}
\affiliation{CAS-Alibaba Quantum Computing Laboratory, Shanghai, 201315, China}

\author{Xiang-Pei Liu}
\affiliation{Shanghai Branch, National Laboratory for Physical Sciences at Microscale and Department of Modern Physics, University of Science and Technology of China, Hefei, Anhui 230026, China}
\affiliation{CAS Center for Excellence and Synergetic Innovation Center in Quantum Information and Quantum Physics, University of Science and Technology of China, Shanghai, 201315, China}
\affiliation{CAS-Alibaba Quantum Computing Laboratory, Shanghai, 201315, China}

\author{Xiao-Qiong Wang}
\affiliation{Shanghai Branch, National Laboratory for Physical Sciences at Microscale and Department of Modern Physics, University of Science and Technology of China, Hefei, Anhui 230026, China}
\affiliation{CAS Center for Excellence and Synergetic Innovation Center in Quantum Information and Quantum Physics, University of Science and Technology of China, Shanghai, 201315, China}
\affiliation{CAS-Alibaba Quantum Computing Laboratory, Shanghai, 201315, China}

\author{Yu-Xuan Wang}
\affiliation{Shanghai Branch, National Laboratory for Physical Sciences at Microscale and Department of Modern Physics, University of Science and Technology of China, Hefei, Anhui 230026, China}
\affiliation{CAS Center for Excellence and Synergetic Innovation Center in Quantum Information and Quantum Physics, University of Science and Technology of China, Shanghai, 201315, China}
\affiliation{CAS-Alibaba Quantum Computing Laboratory, Shanghai, 201315, China}

\author{Yu-Ao Chen}
\affiliation{Shanghai Branch, National Laboratory for Physical Sciences at Microscale and Department of Modern Physics, University of Science and Technology of China, Hefei, Anhui 230026, China}
\affiliation{CAS Center for Excellence and Synergetic Innovation Center in Quantum Information and Quantum Physics, University of Science and Technology of China, Shanghai, 201315, China}
\affiliation{CAS-Alibaba Quantum Computing Laboratory, Shanghai, 201315, China}

\author{Jian-Wei Pan}
\affiliation{Shanghai Branch, National Laboratory for Physical Sciences at Microscale and Department of Modern Physics, University of Science and Technology of China, Hefei, Anhui 230026, China}
\affiliation{CAS Center for Excellence and Synergetic Innovation Center in Quantum Information and Quantum Physics, University of Science and Technology of China, Shanghai, 201315, China}
\affiliation{CAS-Alibaba Quantum Computing Laboratory, Shanghai, 201315, China}
\affiliation{Physikalisches Institut, Ruprecht-Karls-Universität Heidelberg, Im Neuenheimer Feld 226, 69120 Heidelberg, Germany}



\begin{abstract}

We use D1 gray molasses to achieve Bose-Einstein condensation of a large number of $^{41}$K atoms in an optical dipole trap. By combining a new configuration of compressed-MOT with D1 gray molasses, we obtain a cold sample of $2.4\times10^9$ atoms with a temperature as low as 42 $\mu$K. After magnetically transferring the atoms into the final glass cell, we perform a two-stage evaporative cooling. A condensate with up to $1.2\times10^6$ atoms in the lowest Zeeman state $|F=1,m_F=1\rangle$ is achieved in the optical dipole trap. Furthermore, we observe two narrow Feshbach resonances in the lowest hyperfine channel, which are in good agreement with theoretical predictions.

\end{abstract}

\maketitle

\date{\today}

\section{INTRODUCTION}

Quantum degenerate gases of neutral atoms not only provide platforms for studying few- or many-body quantum systems~\cite{bloch2008many,bloch2012quantum}, but are also promising candidates for high-precision quantum metrology~\cite{peters2001high} and scalable quantum computation~\cite{trotzky2008time}. In the past few decades, tremendous experimental efforts have been devoted to achieving quantum degenerate gases with alkali~\cite{anderson1995observation}, alkaline-earth~\cite{stellmer2009bose}, and rare-earth metals~\cite{lu2011strongly}. Due to the existence of both fermionic and bosonic isotopes, a dilute gas of potassium is of particular interest for studying Bose-Fermi and Fermi-Fermi mixtures with tunable interactions. For instance, a mixture of $^{40}$K and $^{41}$K represents an ideal test bed for exploring impurity problems~\cite{schirotzek2009observation} and strongly interacting Bose-Fermi mixtures~\cite{wu2011strongly}.

\begin{figure}[htbp]
\includegraphics[width=\columnwidth]{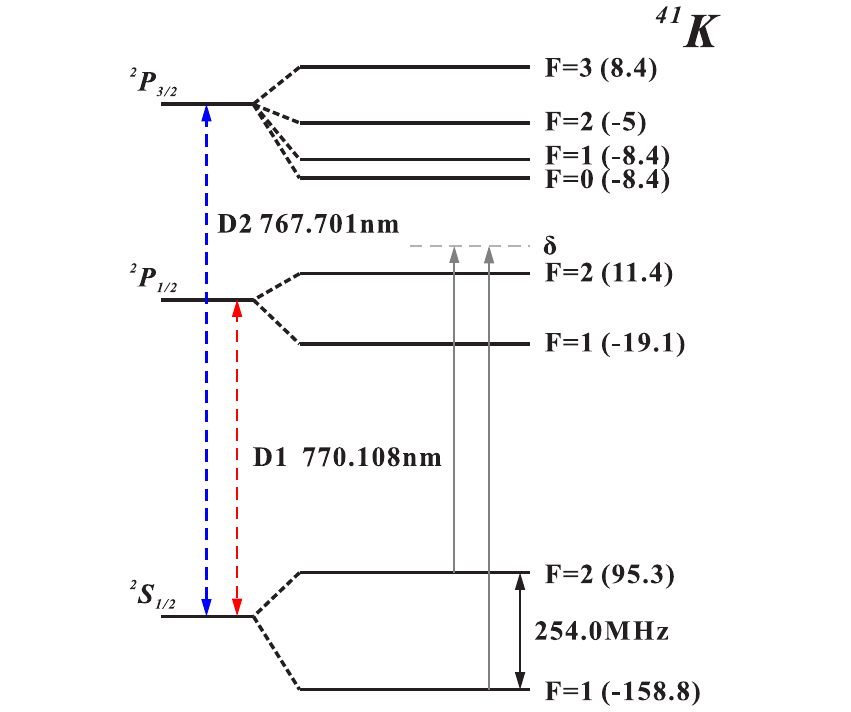}
\caption{(Color Online) The energy-level diagram of $^{41}$K atoms. The natural linewidth of the D2-transition of $^{41}$K is 6 MHz, which is comparable to its excited hyperfine splitting. The laser detuning for gray molasses is defined as $\delta$ for both cooling and repumping transitions.}
\label{Fig1}
\end{figure}

A Bose-Einstein condensate (BEC) of $^{41}$K was first achieved by sympathetic cooling with $^{87}$Rb~\cite{modugno2001bose} in 2001. With a positive background scattering length, $^{41}$K could be directly evaporatively cooled into a Bose-Einstein condensate~\cite{falke2008potassium,kishimoto2009direct}. Moreover, $^{41}$K has proven to be an efficient coolant for $^{6}$Li and $^{40}$K, due to their favorable interspecies background scattering lengths~\cite{tiemann2009coupled,wu2011strongly}. For most alkali-metal atoms, the temperature can be easily reduced below the Doppler limit through Sisyphus cooling in optical molasses. However, the D2 excited hyperfine splitting of $^{41}$K is only 17 MHz, which is unresolved in comparison to its natural linewidth ($\Gamma$=6 MHz) (see Fig.~\ref{Fig1}). Therefore, conventional sub-Doppler cooling techniques are ineffective, making direct evaporative cooling of $^{41}$K an arduous task. In order to overcome this obstacle, different laser-cooling techniques have been developed in recent years~\cite{landini2011sub,mckay2011low}. For example, using the narrower 4s-5p transition of potassium, temperatures around 60 $\mu$K have been observed, equivalent to half of the D2 transition Doppler-cooling limit~\cite{mckay2011low}. However, additional laser sources at 405 nm and optical setup are required for the implementation of this scheme, increasing its experimental complexity. Recently, a so-called gray molasses sub-Doppler cooling technique has been intensively exploited; this method takes advantage of Sisyphus cooling~\cite{dalibard1989laser} and velocity selective coherent population trapping (VSCPT)~\cite{aspect1988laser} to cool atoms well below the Doppler limit. To date, gray molasses has become a well-established method for achieving sub-Doppler cooling of $^{6}$Li~\cite{sievers2015simultaneous,burchianti2014efficient}, $^{7}$Li~\cite{grier2013lambda}, $^{23}$Na~\cite{PhysRevA.93.023421}, $^{39}$K~\cite{nath2013quantum,salomon2014gray}, and $^{40}$K~\cite{sievers2015simultaneous,fernandes2012sub}; however, to the best of our knowledge, the application of the gray molasses technique to $^{41}$K has yet to be reported.

In this letter, we report sub-Doppler cooling of $^{41}$K by using D1 gray molasses. This technique allows us to lower the initial temperature of $2.4\times10^9$ atoms from 418 $\mu$K to 42 $\mu$K. With the help of two-stage evaporative cooling, we are able to produce a $^{41}$K BEC of up to $1.2\times10^6$ atoms in optical dipole trap without a discernible thermal fraction. Furthermore, we investigate the collisional properties of $^{41}$K theoretically and experimentally.

\section{EXPERIMENTAL SET-UP}

\begin{figure}[htbp]
\includegraphics[width=\columnwidth]{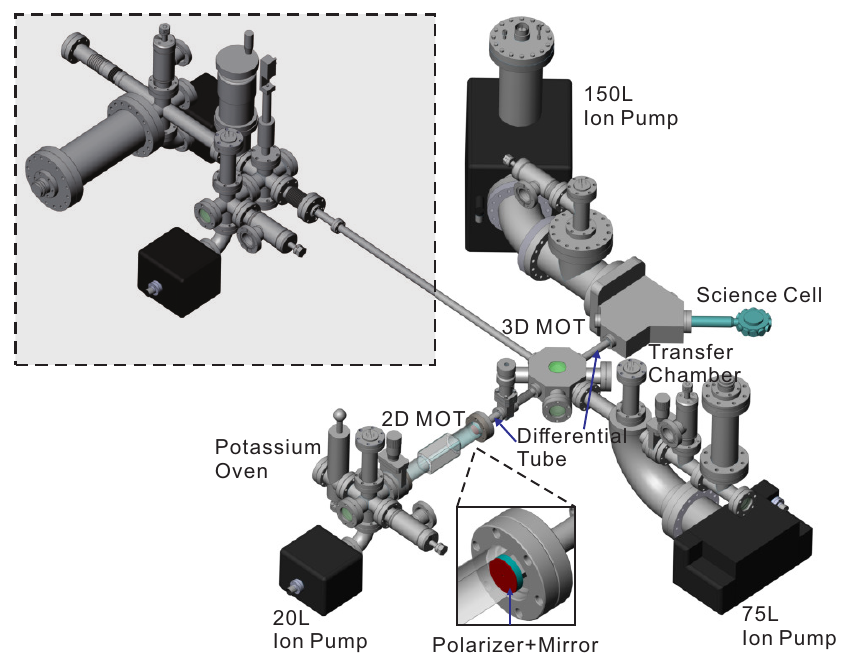}
\caption{(Color Online) Schematics of the vacuum assembly. The shadowing represents the vacuum manifold for future double-species experiments. In the inset, we zoom-in on the 2D MOT chamber to illustrate the optics for our longitudinal molasses. The red and blue optics represent the absorptive polarizer and mirror, respectively.}
\label{Fig2}
\end{figure}

The experimental setup for potassium consists of three main parts: a 2D MOT chamber, a 3D MOT chamber, and a science chamber (Fig.~\ref{Fig2}). The shadow region is to be used for double-species experiments in the future. The 2D MOT chamber is a 4.5 cm$\times$4.5 cm$\times$9.5 cm full-glass cell. A customized absorptive-type polarizer and full-reflecting mirror with a hole in the center are glued to the differential tube between the 2D and 3D MOT chambers. Fifteen pairs of coils are applied to transfer the atomic cloud over a distance of 54 cm to the UHV science chamber for better vacuum and optical access. The achieved vacuum pressure in the full-glass dodecagonal cell is $5\times10^{-12}$ mbar, which greatly increases the lifetime of the atoms.

\section{2D$+$ MOT \& 3D MOT}

Our experimental procedure starts with pre-cooling and confinement of a $^{41}$K atomic vapor using an enhanced 2D$+$ MOT. In a standard 2D$+$ MOT, a pair of counter-propagating beams are applied to the longitudinal molasses. The intensity ratio between incident and retro-reflected beams strongly affect the performance of the 2D$+$ MOT. Thanks to the polarizer and mirror inside the chamber (see Fig.~\ref{Fig2}), we can easily tune the intensity ratio by changing the polarization of an incident laser beam. In our case, the intensity ratio is about 4:1, which is optimized for atomic fluorescence signals. Moreover, an additional blue-detuned laser beam with a 1/e$^{2}$ radius of 0.6 mm is applied in the axial direction to push the atoms into the 3D MOT chamber. We find this push beam to be particularly efficient for achieving a further 50\% enhancement of the atomic flux beyond that allowed by the standard 2D$+$ MOT. Special care is required to slightly misalign the push beam with the captured atomic cloud in the 3D MOT. We also install two UV LED arrays around the 2D MOT chamber, which serve to increase the loading efficiency due to light-induced atom desorption~\cite{klempt2006ultraviolet}. In a 3D MOT chamber, the atomic clouds are captured by six independent laser beams with 1/e$^{2}$ radii of 9 mm. The total intensity of each beam is $I\simeq29\ I_{sat}$, where $I_{sat}$=1.75 mW/cm$^2$ is the saturation intensity of the $^{41}$K D2 transition. We can load more than $4\times10^9$ atoms at a temperature of 5.6 mK in 2 s.

\section{GRAY MOLASSES}

\begin{figure*}[htbp]
\includegraphics[width=\textwidth]{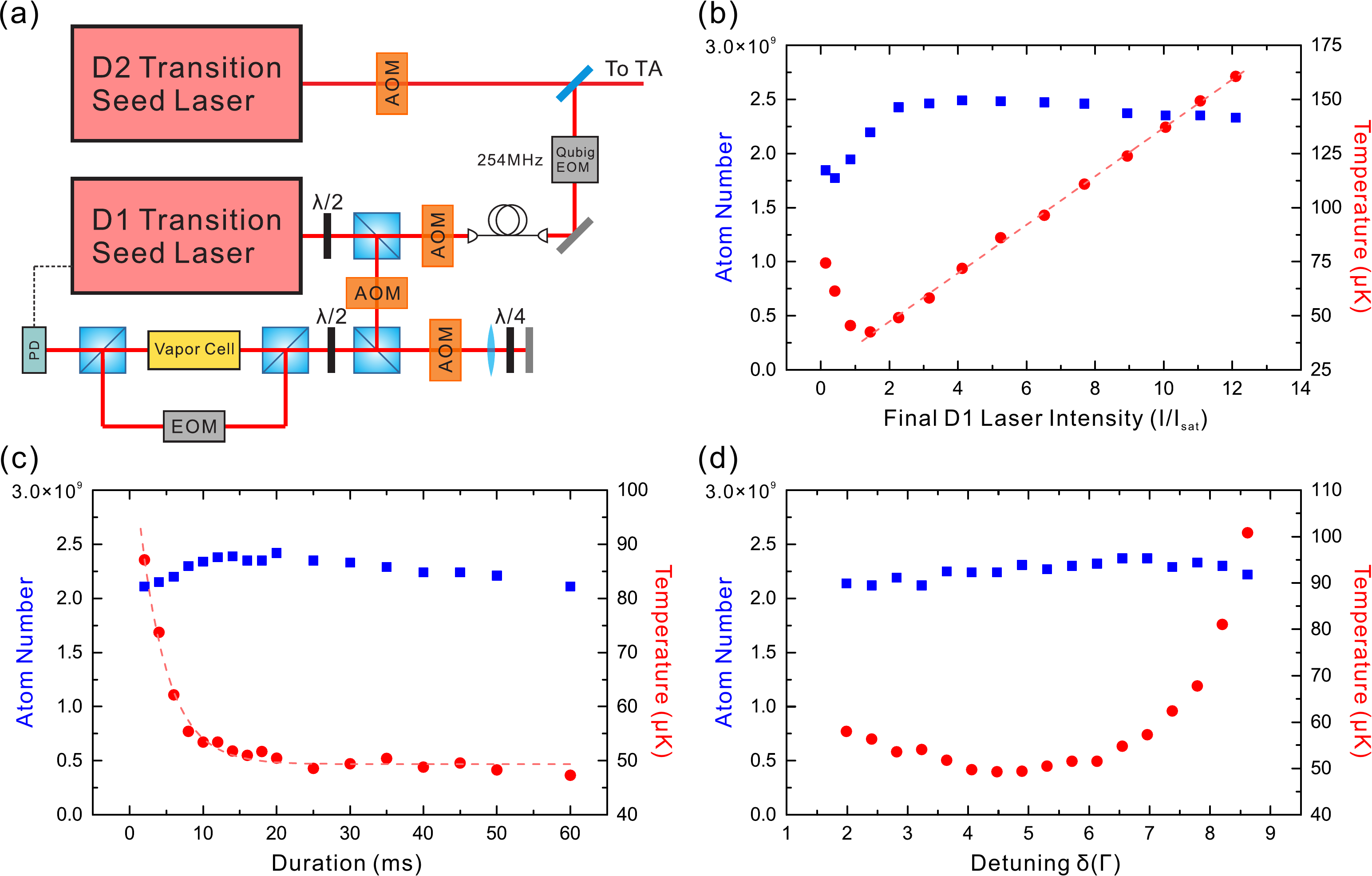}
\caption{(Color Online) (a) Simplified layout of the optical setup for gray molasses. (b) Atom number (blue squares) and temperature (red circles) varying with the final cooling intensity of a D1 laser. The power ratio between the cooler and repumper is fixed at 5:1. Atom temperature scales linearly with the final cooling intensity from $1.5\ I_{sat}$ (42 $\mu$K) to $12\ I_{sat}$ (160 $\mu$K). (c) Atom number (blue squares) and temperature (red circles) as a function of the duration, $\tau_m$. The lowest temperature of 47 $\mu$K is observed with a 1/e cooling time constant of $\tau$=3.84 ms. (d) Atom number (blue squares) and temperature (red circles) as a function of the detuning, $\delta$. Gray Molasses is found to work best for $\delta\in[4\ \Gamma,6\ \Gamma]$.}
\label{Fig3}
\end{figure*}

Gray molasses makes use of both Sisyphus cooling and VSCPT, which is capable of cooling atoms near the single-photon recoil energy. It is composed of two frequencies, which are each blue-detuned by either the $^{41}$K D1 cooling transition $|F=2\rangle\rightarrow |F^\prime=2\rangle$ or the repumping transition $|F=1\rangle\rightarrow |F^\prime=2\rangle$. When the detunings of the two lasers fulfill the Raman condition, a set of long-lived dressed dark states are generated. Dark states and bright states vary throughout real space because of the polarization gradient created by the radiation field. The atoms that are initially in the bright states consume their kinetic energy by climbing the potential hill before being pumped back into dark states, whereas the atoms in dark states that are not sufficiently cold turn into bright states and reenter the cooling circles. This procedure terminates when the atoms in the dark states have a sufficiently long lifetime, leading to a narrow peak in the momentum distribution.

A simplified layout of the D1 laser setup employed for gray molasses is presented in Fig.~\ref{Fig3}(a). A commercial diode laser (Toptica: DL pro) provides the laser at 770 nm, which is locked to the D1 transition of $^{39}$K. The laser passes through a resonant electro-optical modulator (EOM: Qubig EO-K41L3) operating at 254 MHz for generation of the required repumper frequency. The laser beams of the D1 and D2 transitions are superimposed using a customized dichroic mirror before being injected into a tapered amplifier (New Focus: TA 7613-H). Two individual acoustic-optical modulators (AOMs) serve as fast switches, allowing us to rapidly change the output frequency for different cooling phases. With this scheme, the same optical setup is used for both D1 gray molasses and D2 MOT, greatly reducing the experimental complexity.

Before the gray molasses phase, a novel hybrid D1-D2 compressed-MOT (CMOT) technique is implemented to increase the density of the atomic cloud, in a similar manner to that described in G. Salomon's work in Ref~\cite{salomon2014gray}. Compared with a conventional CMOT, the D1 cooling transition $|F=2\rangle\rightarrow |F^\prime=2\rangle$ is used instead of the D2 cooling transition. During this stage, we shut off the input signal of the EOM immediately and ramp up the magnetic gradient from 13 G/cm to 21 G/cm over 10 ms. In order to suppress light-assisted collision, cooling (repumping) intensity is linearly reduced from $11.4\ I_{sat}$ ($1\ I_{sat}$) to $0.1\ I_{sat}$ ($0.5\ I_{sat}$) and detuning is increased from $3.6\ \Gamma$ (-$5.6\ \Gamma$) to $4.5\ \Gamma$ (-$8.4\ \Gamma$). During this process, the RMS radius of the cloud is greatly reduced, yielding an eightfold increase in the peak density. Meanwhile, atom temperature is lowered from 5.6 mK to 418 $\mu$K.

\begin{table*}[htbp]
\begin{tabular*}{\textwidth}{@{\extracolsep\fill}ccccccccccccc}
\toprule
Sequence & $\delta_{2c}$ & $I_{2c}$ & $\delta_{2r}$ & $I_{2r}$ & $\delta_{1c}$ & $I_{1c}$ & $\delta_{1r}$ & $I_{1r}$ & B(G/cm) & T($\mu$K) & N & $\rho$ \\\hline
Loading (2 s) & -8.9 & 11.5 & -5 & 17.5 & - & - & - & - & 13 & 5600 & $4\times10^9$ & $9\times10^{-10}$\\
D1-D2 CMOT (10 ms) & - & - & -5.6$\rightarrow$-8.4 & 1$\rightarrow$0.5 & 3.6$\rightarrow$4.5 & 11.4$\rightarrow$0.1 & - & - & 13$\rightarrow$21 & 418 & $2.6\times10^9$ & $9.4\times10^{-8}$\\
Molasses Holding (2 ms) & - & - & - & - & 5.3 & 12.5 & 5.3 & 2.5 & - & - & - & -\\
Molasses Ramping (15 ms) & - & - & - & - & 5.3 & 12.5$\rightarrow$1.5 & 5.3 & 2.5$\rightarrow$0.3 & - & 42 & $2.4\times10^9$ & $5.4\times10^{-6}$\\
\toprule
\end{tabular*}
\caption{Optimal experimental parameters before the optical pumping stage. The unit of $\delta$ is the natural linewidth of the D2 transition of $^{41}$K ($\Gamma$=6 MHz) and the unit of I is its saturation intensity ($I_{sat}$=1.75 mW/cm$^2$). The subscripts 1 and 2 refer to D1 and D2 transitions, while c and r refer to cooling and repumping, respectively. We also present the time sequence, magnetic field gradient B, temperature T, atom number N, and phase-space density $\rho$ at each stage.}
\label{tab.1}
\end{table*}

In the gray molasses stage, the D2 repumping laser and magnetic field are switched off while the input signal of the EOM is switched on. In the following analysis, we use the same global detuning, $\delta$, for both D1 cooling and repumping transitions to fulfill the Raman condition shown in Fig.~\ref{Fig1}. As the capture efficiency is strongly affected by the laser intensity~\cite{sievers2015simultaneous}, we set the laser power to its maximum at the beginning of gray molasses phase, yielding a temperature of 160 $\mu$K. With an initial atom number of $2.6\times10^9$ in the CMOT, a maximum of $2.4\times10^9$ atoms can be captured by the gray molasses, corresponding to a capture efficiency of 92\%. We observe a further cooling of atoms when the laser intensity is ramped down to a lower value. Moreover, the lowered intensity is still capable of capturing almost all of the atoms in the gray molasses. Thus, we first measure the atom number and temperature of gray molasses by varying the final the D1 laser intensity. The loading time is fixed at 2 ms as gray molasses reaches its maximum capture rate, and then the laser intensity is linearly ramped down over the following 15 ms. The cooling/repumping intensity ratio is fixed at a constant ($\sim 5:1$) and the global detuning, $\delta$, is $4\ \Gamma$. As shown in Fig.~\ref{Fig3}(b), the atom temperature scales linearly with the final cooling intensity from $1.5\ I_{sat}$ to $12\ I_{sat}$. We observe a minimum temperature (42 $\mu$K) that occurs at $I_{cool}=1.5\ I_{sat}$. Both atom loss and inefficient cooling occur when the final intensity is below this value. Thus, we make use of this piecewise time sequence in our following measurements.

Then, we study the efficiency of gray molasses as a function of its duration, $\tau_m$, see Fig.~\ref{Fig3}(c). Though the initial atom temperature is 418 $\mu$K, it rapidly decreases over the first 10 ms and reaches an asymptotic temperature of 47 $\mu$K with a 1/e cooling time constant of $\tau$=3.84 ms.

Finally, we study the influence of global detuning, $\delta$, which is presented in Fig.~\ref{Fig3}(d). From the data, we observe that the atom number shows a weak dependence upon the global detuning from $2\ \Gamma$ to $6\ \Gamma$. For $\delta\in[2\ \Gamma,4\ \Gamma]$, the temperature decreases from 60 $\mu$K to 49 $\mu$K. Then, the temperature remains minimal for $\delta\in[4\ \Gamma,6\ \Gamma]$ and increases rapidly when $\delta\geq6\ \Gamma$.

A typical temperature after the gray molasses phase is 42 $\mu$K, and the obtained phase-space density is $5.4\times10^{-6}$. The optimized parameters before the optical pumping stage and the time sequence are presented in Table.~\ref{tab.1}.

\section{MAGNETIC TRANSPORT}

At the end of the gray molasses phase, atoms are depolarized and randomly distributed into hyperfine ground states. Hence, optical pumping is required to prepare atoms in the low-field seeking state $|F=2,m_F=2\rangle$ for magnetic trapping. In order to minimize the absorption-emission cycles during optical pumping, we use D1 line optical pumping, which causes the target state behave like a dark state in the presence of a bias magnetic field. Furthermore, we apply a pair of balanced retro-reflecting optical pumping beams to decrease the displacement of the atoms. With an optimized optical pumping phase, we achieve a loading efficiency of 70\% for the magnetic trap. Then, we adiabatically transfer the atoms a distance of 54 cm over 3 s from the MOT chamber to the science chamber. The magnetic transport consists of fifteen pairs of overlapping quadrupole coils which generate a moving trapping potential by applying time-varying currents. The transfer path has an angle of 45$^{\circ}$, such that the science cell is out of the line-of-sight of the potassium oven (see Fig.~\ref{Fig2}). We finally obtain $7\times10^8$ atoms with a temperature of 200 $\mu$K at a 115 G/cm magnetic gradient in the glass cell.

\section{EVAPORATIVE COOLING}

Although a $^{41}$K BEC has already been achieved in the magnetic trap, the fact that only the low-field seeking state can be magnetically confined limits its further applications. In contrast to magnetic traps, the trapping potential of optical traps is independent of magnetic substates, making the latter well-suited to investigating of spin-mixture~\cite{o2002observation,jochim2003bose} and Feshbach resonance~\cite{chin2010feshbach}. Therefore, it is desirable to achieve a $^{41}$K BEC in the optical dipole trap. In our experiment, due to heating and atom loss in the magnetic transport phase, directly loading atoms into an optical dipole trap is inefficient. To solve this problem, we adopt a two-stage evaporative cooling strategy. Atoms are first evaporatively cooled in the optically plugged magnetic trap, and then transferred into the optical dipole trap for subsequent evaporation.

\begin{figure}[htbp]
\includegraphics[width=\columnwidth]{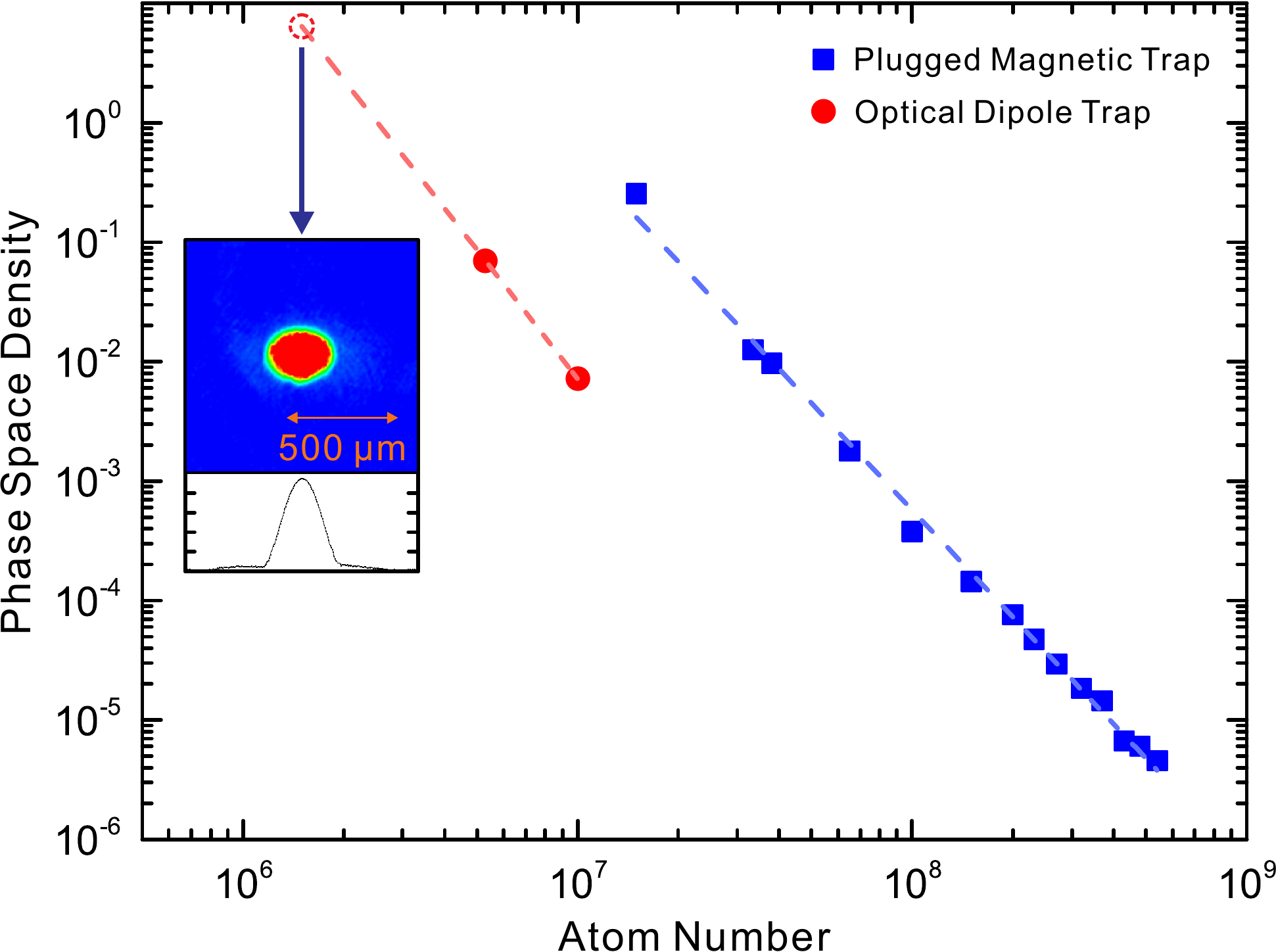}
\caption{(Color Online) Phase-space density as a function of atom number during two-stage evaporative cooling. The blue squares (red circles) represent evaporation in the optically plugged magnetic trap (optical dipole trap). The inset shows the false-color absorption image and the 1D-integrated density profile of a BEC after a 25 ms time-of-flight.}
\label{Fig4}
\end{figure}

In order to avoid Majorana spin-flip, a tightly focused 532 nm laser is used to plug the center of the quadrupole trap. The laser power is 11 W with a 1/e$^{2}$ radius of 34 $\mu $m, which provides a potential barrier of 360 $\mu$K. Evaporation is performed by driving the $|F=2,m_F=2\rangle\rightarrow|F=1,m_F=1\rangle$ transition, which is 254 MHz under zero magnetic field. In the first trail, we ramp up the magnetic gradient from 115 G/cm to 198 G/cm over 200 ms to increase the collision rate and then scan the RF-knife from 340 MHz to 265 MHz linearly over 11 s. In the second trail, in order to suppress three-body loss, we decompress the trap back to 110 G/cm over 160 ms. Afterwards, the RF-knife is further reduced from 261.5 MHz to 257 MHz over 2 s. Finally, we obtain a cloud of $1.5\times10^7$ atoms at 6.5 $\mu$K with an achieved phase-space density of 0.26. The evolution of phase-space density as a function of atom number is plotted in Fig.~\ref{Fig4}. The total efficiency, $\Gamma$, of evaporative cooling is $2.97\pm0.05$, where $\Gamma=-d(lnPSD)/d(lnN)$.

After evaporative cooling is performed in the optically plugged magnetic trap, we transfer the cold sample into a single beam dipole trap. Our trapping potential is generated by a 170 mW, 1064 nm laser with a single spatial and longitudinal mode. The beam is focused into a 1/e$^{2}$ radius of 24 $\mu$m, which corresponds to a 23 $\mu$K trap depth for $^{41}$K. Experimentally, we adiabatically ramp down the magnetic gradient and increase the laser intensity simultaneously over 100 ms. With optimized parameters, the total atom number transferred into the optical trap is $1.2\times10^7$ at 12.2 $\mu$K. Then, we immediately transfer atoms into the hyperfine ground state $|F=1,m_F=1\rangle$ by performing a Landau-Zener sweep. The forced evaporation is accomplished by exponentially decreasing the laser power by a factor of 24 in 3 s. Additional axial confinement is provided by the magnetic curvature under a 300 G bias field, which becomes dominant near the end of evaporation. Finally, we observe a $^{41}$K BEC of up to $1.2\times10^6$ atoms without a discernible thermal fraction (see the inset of Fig.~\ref{Fig4}). The calculated axial and radial trapping frequencies are about $2\pi\times3.7$ Hz and $2\pi\times190$ Hz, respectively.

\section{DETECTION OF FESHBACH RESONANCE}

Feshbach resonance is an essential ingredient for controlling interatomic interactions. Different theoretical approaches for predicting two-body collisional behaviors have been developed in recent years and shown good agreement with experiment results~\cite{stoof1988spin,wille2008exploring,houbiers1998elastic,hanna2009prediction}. Coupled-Channel calculation is a numerical solution method of the coupled Schr\"{o}dinger equations which can directly obtain the scattering length as a function of the bias field~\cite{stoof1988spin}.

\begin{figure}[htbp]
\includegraphics[width=\columnwidth]{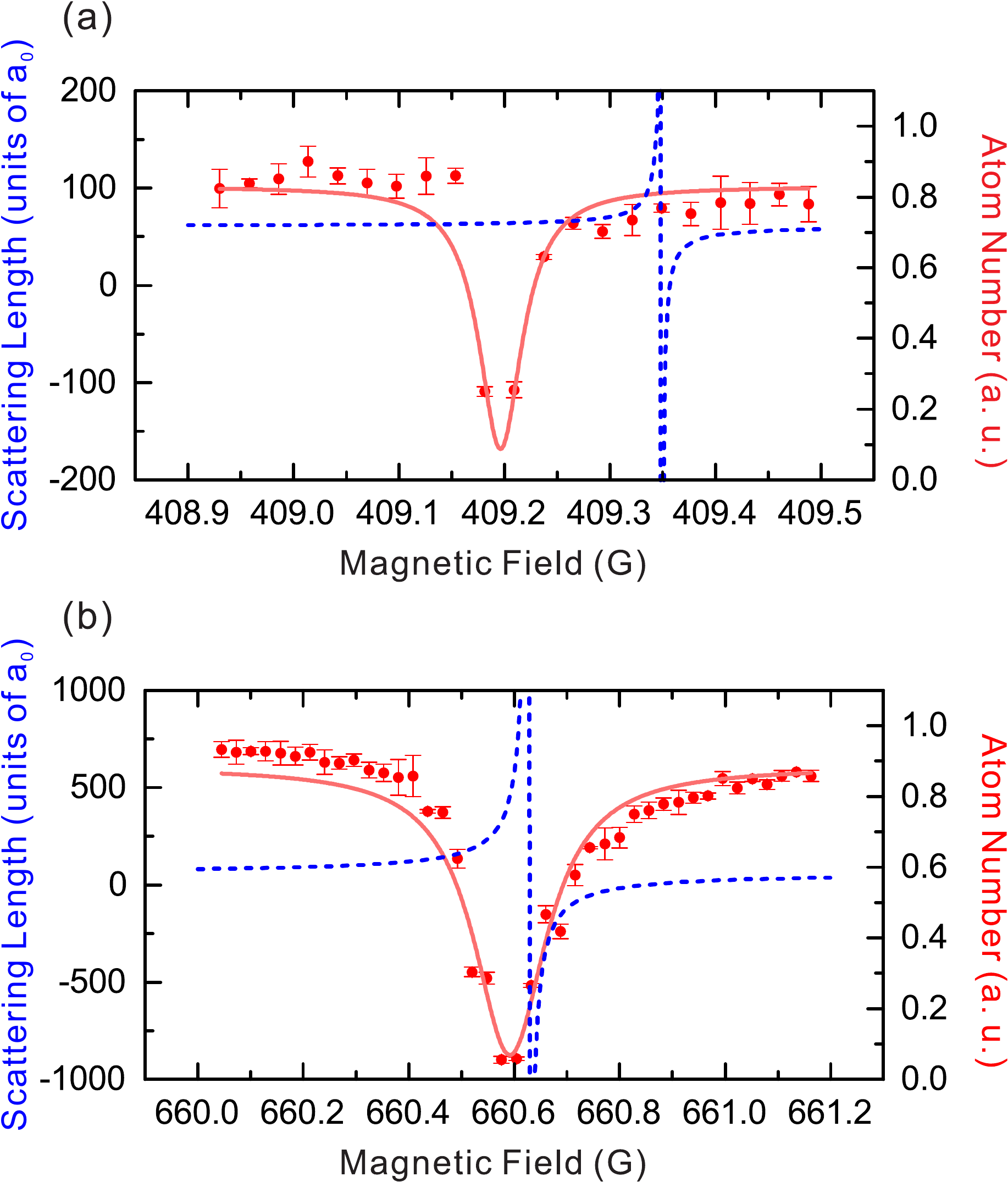}
\caption{(Color Online) Inelastic loss spectra in the lowest hyperfine channel of $^{41}$K. The measurement data (red circles) are fitted by Lorentz functions. The resonance points are located at 409.2 G (a) and 660.6 G (b), respectively. The blue dashed lines represent the theoretical predictions of scattering length in the units of Bohr radius according to a Coupled-Channel calculation; these predictions are in good agreement with the experimental results.}
\label{Fig5}
\end{figure}

Here, s-wave Feshbach resonances of $^{41}$K in the state $|F=1,m_F=1\rangle$ are theoretically simulated using a Coupled-Channel calculation, based on well-established singlet $X^1\Sigma$ and triplet $a^3\Sigma$ potentials~\cite{falke2008potassium}. The theoretical predictions of resonance locations (409.3 G \& 660.6 G) serve as a guide for the experiment. We perform inelastic loss spectroscopy to detect Feshbach resonances where enhanced atom losses occur due to three-body decay. In the following measurements, the initial status of $^{41}$K is $N=3.3\times10^6$ and $T=300$ nK. We first ramp the magnetic field to the desired value and then hold it there for a period of time before measuring the residual atom number in the trap. In order to avoid measurement fluctuations during fast deactivation of the magnetic field, the atom number is directly measured by high-field absorption imaging. Reproducible loss features are observed around 409.2 G and 660.6 G, as shown in Fig.~\ref{Fig5}; the blue dashed lines are theoretical predictions of the scattering length based on the Coupled-Channel calculation. The inelastic loss spectra are fitted with Lorentz functions, which satisfactorily agree with simulation results. The accuracy of the magnetic field is calibrated by RF spectroscopy between the two lowest hyperfine states of $^{41}$K for several points between 400 G and 700 G, and found to be better than 50 mG.

\section{CONCLUSION}

In summary, we have achieved a large production of a $^{41}$K BEC in an optical dipole trap. Our method is based on the a combination of enhanced 2D$+$ MOT, D1-D2 CMOT, D1 gray molasses, and a two-stage evaporation strategy. D1 gray molasses provides an effective sub-Doppler cooling mechanism for $^{41}$K. We make use of two successive cooling phases, yielding a high capture rate (92\%) as well as a low temperature (42 $\mu$K). After magnetically transferring the cloud into the science chamber, evaporative cooling is first performed in an optically plugged magnetic trap with an overall efficiency of $2.97\pm0.05$. This procedure allows us to enhance the phase-space density of atoms by more than four orders of magnitude. The subsequent evaporative cooling is performed in a single beam optical dipole trap, producing a pure BEC of more than $1.2\times10^6$ atoms, which is four times larger than that in any previous experiment~\cite{kishimoto2009direct}. The collisional properties of $^{41}$K in the state $|F=1,m_F=1\rangle$ are studied both theoretically and experimentally. Two resonance positions are observed at 409.2 G and 660.6 G, as has been predicted theoretically. These results set a new benchmark for generation of a BEC with $^{41}$K atoms.

We thank useful discussions with Jue Nan for Coupled-Channel calculations. This work has been supported by the NSFC of China, the CAS, and the National Fundamental Research Program (under Grant No. 2013CB922001). X.-C. Yao acknowledges support from the Alexander von Humboldt Foundation.

\end{document}